\documentclass[prd,a4paper,showpacs,preprint]{revtex4}%
\usepackage{amsmath}
\usepackage{amsfonts}
\usepackage{amssymb}
\usepackage{graphicx}%
\providecommand{\U}[1]{\protect\rule{.1in}{.1in}}

\begin{document}
\title{Classical tests of General Relativity in thick branes}
\author{F. Dahia}
\affiliation{Dep. of Physics, Univ. Fed. da Para\'{\i}ba, Jo\~{a}o Pessoa, Para\'{\i}ba, Brazil}
\author{Alex de Albuquerque Silva}
\affiliation{Dep. of Physics, Univ. Fed. de Campina Grande, Sum\'{e}, Para\'{\i}ba, Brazil.}
\pacs {11.10.Kk, 04.50.-h, 04.25.-g,04.80.-y  }

\begin{abstract}
Classical tests of General Relativity in braneworld scenarios have been
investigated recently with the purpose of posing observational constraints on
parameters of some models of infinitely thin brane. Here we consider the
motion of test particles in a thick brane scenario that corresponds to a
regularized version of the Garriga-Tanaka solution, which describes a black
hole solution in RSII model, in the weak field regime. By adapting a mechanism
previously formulated in order to describe the confinement of massive tests
particles in a domain wall (that simulates classically the trapping of the
Dirac field in a domain wall), we study the influence of the brane thickness
on the four-dimensional (4D) path of massless particles. Although the geometry
is not warped and, therefore, the bound motion in the transverse direction is
not decoupled from the movement in the 4D-world, we can find an explicit
solution for the light deflection and the time delay, if the motion in the
fifth direction is a high frequency oscillation. We verify that, owing to the
transverse motion, the light deflection and the time delay depend on the
energy of the light rays. This feature may lead to the phenomenon of
gravitational rainbow. We also consider the problem from a semi-classical
perspective, investigating the effects of the brane thickness on the motion of
the zero-mode in the 4D-world.

\end{abstract}
\maketitle

\section{Introduction}

Recently we have seen a renewed interest in higher dimensional theories mainly
motivated by braneworld models that were originally considered a promissing
theoretical framework for solving the hierarchy problem \cite{witten,brane}.
In braneworld models, our ordinary four-dimensional space-time is viewed as a
submanifold isometrically embedded in an ambient space of higher dimensions.
An essential feature in braneworld scenarios is the confinement of matter and
fields in the brane, while gravity may propagate in all dimensions. As a
consequence, the extra dimensions can be much larger than the Planck length
$-$ the compactification scale of the original Kaluza-Klein theory (the first
modern extra-dimensional theory) $-$ without introducing irremediable
conflicts between theory and the current observational data \cite{brane}. As a
matter of fact, the so-called RSII model \cite{rs2}, in which the brane is
embedded in five-dimensional space endowed with a negative cosmological
constant, is a phenomenologically consistent model despite the extra dimension
is non-compact. Indeed, the existence of a localized zero-mode of the
perturbation of the metric ensures that the gravitational field reproduces the
four-dimensional behavior for large distances \cite{rs2}.

As gravity has access to extra dimensions, then experimental tests involving
the gravitational field are invaluable tools for investigating the existence
of hidden dimensions in braneworld theories. Therefore, the study of black
holes in higher-dimension theories (for a review, see \cite{BH}) can play a
significant role in revealing traces of the hidden dimensions. However, in the
RSII model specifically, the task of finding an exact solution that represents
a realistic black hole confined in the brane proved to be very difficult
\cite{BH}. Among several attempts to find a static black hole solution in RSII
model, we would like to highlight here the Garriga-Tanaka solution
\cite{garriga}, which is expected to be the weak field regime of a black hole
in RSII model, and the exact solutions known as DMPR and CFM \cite{dmpr,cfm},
which represent acceptable static and spherically symmetric geometries in the
brane. More recently, it was claimed that an exact and complete solution of a
realistic black hole in the RSII model was finally obtained \cite{wise}.

It follows naturally from these studies the interest in investigating the
so-called classical tests of General Relativity in this scenario. There are
many works concerned with the problem of posing experimental constraints on
the parameters of those models, through the study of the influence of the
hidden dimensions over the motion of test particles in the 4D-world
\cite{tests KK,tests brane,keeton}. In particular, the light deflection and
time delay in the infinitely thin brane of the RSII model is discussed in Ref.
\cite{keeton}. However, the classical tests impose weaker bounds than those
obtained in laboratory tests of the square inverse law \cite{newton}, for
instance. Nevertheless, as the constraints are obtained independently, by
exploring different physical aspects of the system, the classical tests should
be considered as important complementary tests, at least.

String theory is the original inspiration for the braneworld models and, in
this context, the brane is infinitely thin \cite{witten}. However, in the
field-theory framework, the brane in the RSII model, for instance, can be
described by a scalar field which is found in a domain wall configuration
\cite{rubakov}. In this case, the matter can be trapped, by means of
Yukawa-type interaction with the scalar field, in the core of the brane, which
now has a non-null thickness \cite{rubakov}. The thick brane solutions are
considered as regularized versions of the infinitely thin brane model, since
they recover the corresponding thin solution in the limit when the thickness
goes to zero. There are many studies on the effects of the brane thickness
over physical systems \cite{thick}. However, as far as we know, no exact
solution of a confined black hole in a thick 3-brane was found yet, although
solutions for lower dimension (2-brane) are known \cite{emparan}. Here we want
to discuss an approximate solution of a black hole in the context of thick
brane version of the RSII model. The idea is to propagate the metric defined
in the core of the brane towards the bulk by using the Einstein equations
coupled to a scalar field in the case of a static and axial symmetry, as
outlined in ref. \cite{dahia2}. This method is detailed in section II, where
we find explicitly the metric in terms of a power series with respect to the
transverse coordinate. The solution is valid in the vicinity of the brane
center and far from the black hole.

In section III, we proceed to the analysis of the classical motion of the
particle in the thick brane. However, to describe its motion, first we have to
implement a confinement mechanism for test particles that simulates
classically the localization of the Dirac field in a domain wall. In Ref.
\cite{dahia1} we have proposed a confinement mechanism for massive particle,
which consists in modifying the Lagrangian of the particle by introducing a
coupling with the scalar field. The role of the scalar field is to increase
the effective mass of the particle when its movement has a transverse
component. Based on that ideas, we have developed a formalism in order to
study the confinement of a massless particle in the domain wall with the
particular purpose of discussing the deflection of light and the time delay
caused by a mass $M$ in the thick brane. Despite the fact that metric is not
warped and, therefore, the transversal and the radial motions are not
decoupled, we find explicitly the influence of the extra dimension over the
four-dimensional motion and, as we shall see, the results (regarding the light
deflection and time delay) depend on the energy of the light rays. An
interesting consequence of this dependence is the formation of a gravitational
rainbow when a beam of white light is deviated by a massive body.

In section IV, we investigate this problem from a semi-classical perspective,
considering the quantization of the motion in the fifth direction. We study
the classical motion of the so-called zero-mode (the massless bound state) in
the 4D-world. In particular, we have determined the effect of the brane
thickness on the bending of ligth rays by a body of mass $M$ and, as we shall
see, the deflection angle depends on the width in the extra-dimension
direction of the wave function that describes the zero-mode state.

\section{Gravity in thick branes}

A non-rotating mass $M$ (of a black hole or a star) localized in a brane
should give rise to an axisymmetric, static spacetime in five dimensions. In
such spaces, as is well known, there are coordinates in which the metric
assumes the Weyl canonical form \cite{ruth}. By means of a convenient
coordinate transformation, the metric can be put into a Gaussian form adapted
to the brane:%
\begin{equation}
ds^{2}=-e^{2A(r,z)}dt^{2}+e^{2B(r,z)}dr^{2}+e^{2C\left(  r,z\right)  }%
d\Omega^{2}+dz^{2}%
\end{equation}
where $z=0$ gives the localization of the brane.\qquad\qquad\qquad

In the context of thick brane scenarios in five dimensions, the brane is
usually described as a domain wall generated by a certain scalar field $\phi$.
It is reasonable to expect that a body of mass $M$ will affect the domain wall
solution. However, considering the symmetry of the problem, we might assume
that the new solution will depend only on coordinates $r$ and $z$, i.e.,
$\phi=\phi\left(  r,z\right)  $.

Even in the absence of $M$, the Einstein Equations coupled to the scalar field
($G_{\mu\nu}=\kappa T_{\mu\nu}^{\left(  \phi\right)  }$ and $\square
\phi-V^{\prime}\left(  \phi\right)  =0$, where $\kappa$ is the gravitational
constant in five dimensions, $T_{\mu\nu}^{\left(  \phi\right)  }$ is the usual
energy-momentum tensor of the scalar field and $V^{\prime}\left(  \phi\right)
$ is the derivative of the potential $V\left(  \phi\right)  $ with respect to
the scalar field) are not easy to solve. However, when the potential is
conveniently choosen, an exact solution of a self-gravitating domain wall can
be obtained. For example, taking $V\left(  \phi\right)  =\frac{\lambda}%
{4}\left(  \phi^{2}-\eta^{2}\right)  ^{2}-\frac{\beta\lambda}{3\eta^{2}}%
\phi^{2}\left(  \phi^{2}-3\eta^{2}\right)  ^{2}$, the solution is
\cite{thicksolution}:%
\begin{align}
ds^{2}  &  =e^{2a\left(  z\right)  }\left(  -dt^{2}+dr^{2}+r^{2}d\Omega
^{2}\right)  +dz^{2},\label{metric0}\\
2a\left(  z\right)   &  =-2\beta\ln\left(  \cosh^{2}\left(  \frac
{z}{\varepsilon}\right)  \right)  -\beta\tanh^{2}\left(  \frac{z}{\varepsilon
}\right)  ,\\
\phi &  =\eta\tanh\left(  \frac{z}{\varepsilon}\right)  ,
\end{align}
where $\varepsilon^{2}=2/\lambda\eta^{2}$, $\beta=\kappa\eta^{2}/9$.

This solution can be interpreted as a regularized version of the RSII brane
model. Indeed, in the limit when the parameter $\varepsilon$ (the thickness of
the wall) goes to zero with the condition $2\beta/\varepsilon\equiv\ell
=const$, the RSII solution is recovered, with $\ell$ playing the role of the
curvature radius of $AdS_{5}$ space.

When the mass $M$ is taken into account then the technical difficulty to solve
the equations increases significantly and, as far as we know, no exact
solution is known. Considering this, we have to resort to perturbative methods
in order to approach the problem \cite{dahia2}. The presence of $M$ certainly
modifies both the original metric and the scalar field. At large distances,
where the weak field regime is valid, the modification can be treated as a
small perturbation of the original solution. In this case, we can write:%
\begin{align}
ds^{2}  &  =e^{2a\left(  z\right)  }\left[  -(1+f)dt^{2}+\left(  1+m\right)
d\xi^{2}+r^{2}\left(  1+h\right)  d\Omega^{2}\right]  +dz^{2},\nonumber\\
\phi &  =\eta\left[  \tanh\left(  \frac{z}{\varepsilon}\right)  +k\left(
r,z\right)  \right]  ,
\end{align}
where $f,m,h$ and $k$ are function of $r$ and $z$ which measure the small
deviations of the metric components.

It may happen that, due to the perturbation, the coordinate $z$ will not be
adapted to the level surface of the scalar field anymore. For instance, the
center of the domain wall $\left(  \phi=0\right)  $, which originally
coincides with $z=0$, is now given by the equation $z=-\varepsilon k\left(
r,0\right)  $. However, the set of adapted coordinates can be restored by
means of an appropriate transformation \cite{garriga}. For our purpose it is
convenient to use a Gaussian coordinate system adapted to the center of domain
wall, because, as we shall see, our approach to determine the solution is
based on the propagation of the metric from the center of the brane to the
bulk by the field equation. In this sense, the metric evaluated in the center
constitute a kind of initial data for the bulk solution. Thus, the great
advantage of working in these coordinates is the fact that initial conditions,
i.e., the value that the functions $f,$ $m,$ $h$ and $k$ and its first
derivatives assume in the center of wall, can be easily established. For
example, as the center corresponds to $z=0$, then, we should have
\begin{equation}
k\left(  r,0\right)  =0.\label{InitialS}%
\end{equation}
Another important condition can be immediately deduced based on the fact that
the metric should be symmetric with respect to the center of the wall. As
$\partial_{z}$ is the normal vector of the hypersurface $z=0$, then, it
follows that, in the center of the domain, we should have:
\begin{equation}
\left.  \frac{\partial f}{\partial z}\right\vert _{z=0}=\left.  \frac{\partial
m}{\partial z}\right\vert _{z=0}=\left.  \frac{\partial h}{\partial
z}\right\vert _{z=0}=0.\label{Dz}%
\end{equation}
These functions must satisfy other conditions that we shall discuss later. But
now let us concentrate our attention on the field equation. In the first
approximation order \cite{gio}, those equations are reduced to the following
set:%
\begin{equation}
^{\left(  1\right)  }R_{t}^{t}\equiv\frac{1}{2}f_{zz}+\frac{1}{2}\left(
f_{rr}+\frac{2}{r}f_{r}\right)  e^{-2a}+\frac{1}{2}a^{\prime}\left(
5f_{z}+m_{z}+2h_{z}\right)  =-\frac{2}{3}\kappa V_{0}^{\prime}k\label{dynf}%
\end{equation}

\begin{align}
^{\left(  1\right)  }R_{r}^{r}\equiv\frac{1}{2}m_{zz}+\left(  \frac{1}%
{2}f_{rr}-\frac{1}{r}m_{r}+h_{rr}+\frac{2}{r}h_{r}\right)   &  e^{-2a}%
\nonumber\\
+\frac{1}{4}a^{\prime}\left(  2f_{z}+10m_{z}+4h_{z}\right)   &  =-\frac{2}%
{3}\kappa V_{0}^{\prime}k\label{dynm}%
\end{align}%
\begin{align}
^{\left(  1\right)  }R_{\theta}^{\theta}=\frac{1}{2}h_{zz}+\frac{1}{4}\left(
2h_{rr}+\frac{4}{r}h_{r}+\frac{2}{r}\left(  f_{r}-m_{r}+2h_{r}\right)
\right)   &  e^{-2a}\nonumber\\
+\frac{1}{2}a^{\prime}\left(  f_{z}+m_{z}+6h_{z}\right)  -\frac{1}{r^{2}%
}e^{-2a}\left(  m-h\right)   &  =-\frac{2}{3}\kappa V_{0}^{\prime
}k\label{dynh}%
\end{align}%
\begin{equation}
^{\left(  1\right)  }R_{rz}\equiv\frac{1}{2}f_{zr}+h_{zr}+\frac{1}{r}\left(
h_{z}-m_{z}\right)  =-\kappa k_{r}\phi_{z}^{\left(  0\right)  }\label{constR}%
\end{equation}%
\begin{align}
^{\left(  1\right)  }G_{zz} &  \equiv-\frac{3}{2}a^{\prime}\left(  f_{z}%
+m_{z}+2h_{z}\right)  -\frac{1}{4}\left(  2f_{rr}+4h_{rr}+\frac{4}{r}\left(
f_{r}-m_{r}+3h_{r}\right)  \right)  e^{-2a}+\nonumber\\
&  +\frac{1}{r^{2}}e^{-2a}\left(  m-h\right)  =-\kappa\left(  \phi
_{z}^{\left(  0\right)  }\,k_{z}-V_{0}^{\prime}k\right)  \label{constG}%
\end{align}%
\begin{align}
^{\left(  1\right)  }\square\phi-V_{0}^{\prime\prime}k &  \equiv
k_{zz}+4a^{\prime}k_{z}+\frac{1}{2}\phi_{0}^{\prime}\left(  f_{z}+m_{z}%
+2h_{z}\right)  +\nonumber\\
&  e^{-2a}\left(  k_{rr}+\frac{2}{r}k_{r}\right)  -V_{0}^{\prime\prime
}k=0,\label{dynk}%
\end{align}
where the quantities carrying the indeces $(0)$ and $(1)$ are calculated in
the zero and in the first approximation order with respect to $GM,$ respectively.

This system can be divided into two classes of equations: dynamic equations
(\ref{dynf}), (\ref{dynm}), (\ref{dynh}) and (\ref{dynk}) - which contain
second derivative of the metric with respect to the transverse coordinate
($z)$ - and constraint equations (\ref{constR}) and (\ref{constG}). It is well
known that, owing to the Bianchi identities, the constraints are propagated by
the dynamical equations. This means that if the constraint equations are
satisfied in a certain hypersurface, the dynamic equations ensure that the
constraint equations will be satisfied in an open set of the manifold in the
vicinity of that hypersurface. Thus a solution of the equations (\ref{constR})
and (\ref{constG}) in the hypersurface $z=0$ can be interpreted as an
admissible initial data for the bulk solution of the dynamic equation.

The dynamic equation can be solved by an iterative method. Isolating the
second derivative of the metric or the scalar field on the left hand side of
the equations, we can calculate them on the hypersurface $z=0$ from the
initial data. Derivating the equations with respect to $z$ and repeating the
same procedure successively, we can evaluate the derivatives of any order of
the unknown functions $f,m,h$ and $k$ at $z=0$. With these derivatives, we
construct a power series in respect of $z$. According to Cauchy-Kowalewsky
theorem, this series converges in some open set and represents the solution of
the equations.

We can use this procedure to find an approximate solution of the problem
investigated here. The first setp is to choose an appropriate set of initial
data. As we have already mentioned, the initial data consist of the functions
$f\left(  r,0\right)  ,m\left(  r,0\right)  ,h\left(  r,0\right)  $ and
$k\left(  r,0\right)  $ and its first derivatives defined on the hypersurface
$z=0$ that satisfy the constraint equations (\ref{constR}) and (\ref{constG}).
Some of the initial conditions have been already determined in eqs
(\ref{InitialS}) and (\ref{Dz}). It is interesting to note that, with these
choices, the constraint equation (\ref{constR}) is automatically satisfied.

The remaining set of the initial conditions, i.e, $f\left(  r,0\right)
,m\left(  r,0\right)  ,h\left(  r,0\right)  $ and $\left.  \frac{\partial
k}{\partial z}\right\vert _{z=0}$, can be determined by using the thin brane
solution as inspiration. According to Garriga and Tanaka \cite{garriga} , in
the limit of thin brane, the metric of a matter distribution with mass $M$
localized in the brane is given, in the first approximation order of $GM$, by:%
\begin{equation}
ds^{2}=-\left(  1-\frac{2GM}{r}-\frac{4GM\ell^{2}}{3r^{3}}\right)
dt^{2}+\left(  1+\frac{2GM}{r}+\frac{2GM\ell^{2}}{r^{3}}\right)  dr^{2}%
+r^{2}d\Omega^{2}. \label{garriga}%
\end{equation}
We have to emphasize that this metric is valid in the brane $\left(
z=0\right)  $ in a region far from the source $\left(  r>>GM\ \right)  $.

In order to obtain a link between our thick brane solution and thin brane
solution (\ref{garriga}), we are going to admit that the induced metric in the
center of the domain wall is isometric to the thin brane's geometry. This
condition implies that:%
\begin{align}
f\left(  r,0\right)   &  =-\left(  \frac{2GM}{r}+\frac{4GM\ell^{2}}{3r^{3}%
}\right)  ,\label{f}\\
m\left(  r,0\right)   &  =\frac{2GM}{r}+\frac{2GM\ell^{2}}{r^{3}},\label{g}\\
h\left(  r,0\right)   &  =0. \label{h}%
\end{align}
With this choices, the constraint equation (\ref{constG}) evaluated in $z=0$
imposes an additional initial condition:%
\begin{equation}
\left.  k_{z}\right\vert _{z=0}=0. \label{Dk}%
\end{equation}
The equations (\ref{InitialS}), (\ref{Dz}), (\ref{f}), (\ref{g}), (\ref{h})
and (\ref{Dk}) yield a set of appropriate initial data. Now the bulk solution
can be determined from the dynamical equations by the iterative method. For
instance, calculating the second derivative of the unkown functions in terms
of the initial data we find directly from the equations (\ref{dynf}),
(\ref{dynm}), (\ref{dynh}) and (\ref{dynk}) that:
\begin{equation}
\left.  f_{zz}\right\vert _{z=0}=-\left(  f_{rr}+\frac{2}{r}f_{r}\right)
_{z=0}=\frac{8GM\ell^{2}}{r^{5}},
\end{equation}

\begin{equation}
\left.  m_{zz}\right\vert _{z=0}=-\left(  f_{rr}-\frac{2}{r}m_{r}\right)
_{z=0}=\frac{4GM\ell^{2}}{r^{5}},
\end{equation}

\begin{equation}
\left.  h_{zz}\right\vert _{z=0}=\left[  \frac{2}{r^{2}}m-\frac{1}{r}\left(
f_{r}-m_{r}\right)  \right]  _{z=0}=-\frac{6GM\ell^{2}}{r^{5}},
\end{equation}

\begin{equation}
\left.  k_{zz}^{\left(  1\right)  }\right\vert _{z=0}=0.
\end{equation}
Therefore, up to the second order in the transverse direction, the metric, in
the region $r>>GM$ and $z<<\varepsilon$, is given by%
\begin{align}
ds  &  =-e^{2a}\left[  1-\frac{2GM}{r}\left(  1+\frac{2\ell^{2}}{3r^{2}}%
-\frac{2\ell^{2}}{r^{4}}z^{2}\right)  \right]  dt^{2}\nonumber\\
&  +e^{2a}\left[  1+\frac{2GM}{r}\left(  1+\frac{\ell^{2}}{r^{2}}+\frac
{\ell^{2}}{r^{4}}z^{2}\right)  \right]  dr^{2}\nonumber\\
&  +e^{2a}r^{2}\left(  1-\frac{3GM\ell^{2}}{r^{5}}z^{2}\right)  d\Omega
^{2}+dz^{2}. \label{metricM}%
\end{align}
The scalar solution is simply $\phi=\eta\tanh\left(  \frac{z}{\varepsilon
}\right)  $ in the same approximation order. As a matter of fact, it is
interesting to emphasize that, admitting those initial data, the scalar field
has no modification in any order of the expansion. We can check this by
noticing that the perturbation of scalar field (the function $k$) couples to
the metric by means of the function $\gamma\equiv f+m+2h$ (which is a kind of
trace of the perturbation of the metric). We can determine an equation for
$\gamma$ by taking the following combination $R_{t}^{t}+R_{r}^{r}+2R_{\theta
}^{\theta}-G_{zz}$ from the field equations. This yields
\begin{equation}
\gamma_{zz}+2a^{\prime}\gamma_{z}=4\kappa\left(  -\frac{1}{3}V_{0}^{\prime
}k^{\left(  1\right)  }-\phi_{z}^{\left(  0\right)  }\,k_{z}^{\left(
1\right)  }\right)  .
\end{equation}
On the other hand, the scalar field equation gives%
\begin{equation}
k_{zz}+4a^{\prime}k_{z}+\frac{1}{2}\phi_{0}^{\prime}\gamma_{z}+e^{-2a}%
\nabla^{2}k-V_{0}^{\prime\prime}k=0.
\end{equation}
By using the initial data, $\gamma_{z}\left(  0,r\right)  =0$, $k\left(
0,r\right)  =0$ and $k_{z}\left(  0,r\right)  =0$ in the above equations, we
can conclude that the derivatives of $\gamma_{z}$ and $k$ of any order are
zero when calculated in $z=0$. Thus, there is no correction for the scalar
field $\phi=\eta\tanh\left(  \frac{z}{\varepsilon}\right)  $ in the first
order of $GM$, admiting those initial conditions.

\section{Motion of massless test particles}

Let us now consider a test particle moving in the five dimensional spacetime
whose metric is given by (\ref{metricM}). It is well known that, in a thick
brane of RSII type, geodesics are not stable, i.e., particles escape to the
bulk if they suffer any transversal perturbation. So, it is necessary to
provide a confinement mechanism for test particles in the brane. In ref.
\cite{dahia1}, we have proposed, based on the Yukawa interaction between
fermions and the domain wall, a Lagrangian to describe the particle's motion
in this context. As it is was shown in that reference, the Lagrangian has the
effect of increasing the effective mass of the particle, due to the
interaction with the scalar field, and this modification ensures the
localization of the particle.

The new Lagrangian is given by
\begin{equation}
L=-\sqrt{m^{2}+h^{2}\varphi^{2}}\sqrt{-\tilde{g}_{AB}\dot{x}^{A}\dot{x}^{B}}%
\end{equation}
where $m$ is the rest mass of the particle and $h$ is the coupling constant of
the interaction.

Calculating the $5D$-momentum $P_{A}$ of the particle, which is obtained by
taking $\partial L/\partial\dot{x}^{A}$, we find, using the condition
$\tilde{g}_{AB}\dot{x}^{A}\dot{x}^{B}=-1,$ that the effective mass of the
particle is influenced by the scalar field according to the expression:%
\begin{equation}
P^{A}P_{A}=-\left(  m^{2}+h^{2}\varphi^{2}\right)  . \label{mass}%
\end{equation}
Of course, the usual relation is recovered turning off the interaction, i.e.,
taking $h=0$. It is worthy of mention that a similar kind of Lagrangian was
also employed, in a different context, to describe the interaction between
test particles and dilatonic fields \cite{kim}.

From the Euler-Lagrange equations, we can express the equation of motion in
the following form:%
\begin{equation}
\overset{\cdot\cdot}{x}^{A}+\Gamma_{BC}^{A}\dot{x}^{B}\dot{x}^{C}=a^{A},
\label{eqmotion}%
\end{equation}
where $\Gamma_{BC}^{A}$ is the Levi-Civita connection and $a^{A}$ is the
proper acceleration of the particle due to its interaction with the scalar
field and it can be written as the gradient of the effective mass of the
particle:%
\begin{equation}
a^{A}=-\Pi_{\quad}^{AC}\tilde{\nabla}_{C}\ln\left(  m^{2}+h^{2}\varphi
^{2}\right)  ,
\end{equation}
where $\Pi^{AC}=\tilde{g}^{AC}+\dot{x}^{A}\dot{x}^{C}$ is the projection
tensor into the four-space orthogonal to the particles' proper velocity
$\dot{x}^{A}$.

The equation of motion (\ref{eqmotion}) cannot be applied to study the
propagation of a masslees particle in the thick brane, because, the proper
acceleration $a^{A}$ is not well defined in the center of brane for $m=0$.
Thus, we need to follow a new procedure to describe the motion of light rays
in this context. A possible way is to start with the dispersion relation
(\ref{mass}). Taking $m=0$ and introducing a new function $G\left(
\varphi\right)  $, we can write:%
\begin{equation}
P^{A}P_{A}=-h^{2}G^{2}\left(  \varphi\right)  . \label{massG}%
\end{equation}
The function $G\left(  \varphi\right)  $ must satisfies the condition
$G\left(  0\right)  =0$, to guarantee that in the brane center the particle's
mass is equal to zero. We could have $G\left(  \varphi\right)  =\varphi$, for
instance. The equation (\ref{massG}) shows that the interaction with the
scalar field generates mass for the particle if it moves in the extra
dimension. As we shall see later, this mechanism is responsible for keeping
the particle confined in the brane provided $G\left(  \varphi\right)  $ is a
function with appropriate properties. On the other hand, if a particle is
moving strictly in the center of the brane, then, the usual relation for a
massless particle $P_{A}P^{A}=0$ is recovered. In this case, that particle
would have no mass from the 4D-viewpoint.

Based on the wave mechanics, which establishes a relation between the linear
momentum $P^{A}$ of a free particle and the wave vector $K^{A}$ of its
corresponding plane wave, we are led to propose that
\begin{equation}
K^{A}K_{A}=-h^{2}G^{2}\left(  \varphi\right)  . \label{dispersion}%
\end{equation}

The equation of motion for a massless particle can be obtained by taking the
covariant derivative of the above equation and admitting that $K_{A}%
=\nabla_{A}S,$ i.e, the wave vector is the gradient of the wave phase $S$ (a
scalar function). Then, by using the condition $\nabla_{B}K_{A}=\nabla
_{A}K_{B},$ it follows that:%
\begin{equation}
K^{A}\nabla_{A}K_{B}=-\frac{1}{2}h^{2}\nabla_{B}G^{2}\left(  \varphi\right)  .
\label{motion}%
\end{equation}
Of course, the particle world-line is obtained by integrating the equation
$dx^{A}/d\lambda=K^{A}$, where $\lambda$ is some affine parameter.

With the help of the equation (\ref{motion}), we now are able to study the
motion of light rays. First, we want to discuss some general aspects of the
motion when $M=0$. In this case, the geometry of the thick brane is described
by the metric (\ref{metric0}). As that metric is warped, the transversal
motion decouples from the movement in the other directions. After some
manipulation, we can show that the evolution of the $z$-coordinate can be
directly integrated and its first-integral can be put in the form:%
\begin{equation}
e^{2a}\dot{z}^{2}=\dot{z}_{0}^{2}-V_{0}\left(  z\right)  , \label{tmotion}%
\end{equation}
where $\dot{z}_{0}$ is a constant related to the initial condition and
$V_{0}\left(  z\right)  \equiv h^{2}e^{2a}G^{2}\left(  \varphi\right)  $ works
as an effective potential for the transversal motion. It is clear that, by an
appropriate choice of $G$, the potential $V_{0}$ is capable of confining the
particle around the brane center as we have already mentioned.

When the mass $M$ is taken into account the spacetime geometry is no longer
described by a warped metric and the decoupling between the motions does not
exist anymore. Considering the perturbation on the geometry caused by the mass
$M$, as described in the metric (\ref{metricM}), the equation of the
transversal motion now assumes the following form:%

\begin{equation}
\frac{d}{d\lambda}\left[  e^{2a}\left(  h^{2}G^{2}\left(  \varphi\right)
+\dot{z}^{2}\right)  \right]  =e^{2a}\dot{z}q_{\mu\nu}\dot{x}^{\mu}\dot
{x}^{\nu}, \label{tmotionM}%
\end{equation}
where%
\begin{equation}
q_{\beta\gamma}dx^{\beta}dx^{\gamma}=e^{2a}\left[  -\left(  \frac{8GM\ell^{2}%
}{r^{5}}z\right)  dt^{2}+\left(  \frac{4GM\ell^{2}}{r^{5}}z\right)
dr^{2}-\left(  \frac{6GM\ell^{2}}{r^{3}}z\right)  d\Omega^{2}\right]  .
\label{q}%
\end{equation}

As we can see, the transverse motion is now coupled to the movement in the
other directions of the brane through the term $q_{\mu\nu}$ which is of the
order of $GM$. Thus, to proceed further, we have to analyze the evolution of
other coordinates. By exploring the axial and temporal symmetries of the
spacetime we get the equations:
\begin{align}
-g_{tt}\dot{t}  &  =E,\label{E}\\
g_{\phi\phi}\dot{\phi}  &  =L, \label{L}%
\end{align}
where, for a convenient choice of the affine parameter, $E$ and $L$ correspond
to the particle's energy and angular momentum respectively, as measured by
asymptotic observers that lie in the center of the brane. Examining the
equation (\ref{eqmotion}) for $x^{A}=\theta$, we can also verify that
$\theta=\pi/2$ is a possible solution. By using these equations and also
(\ref{dispersion}), we can rewrite (\ref{tmotionM}) in this way:
\begin{align}
\frac{d}{d\lambda}  &  \left[  e^{2a}\left(  h^{2}G^{2}\left(  \varphi\right)
+\dot{z}^{2}\right)  \right]  =\nonumber\\
&  \left[  \left(  \frac{q_{tt}}{g_{tt}}-\frac{q_{rr}}{\hat{g}_{rr}}\right)
\frac{E^{2}}{\tilde{g}_{tt}}+\left(  \frac{q_{\phi\phi}}{g_{\phi\phi}}%
-\frac{q_{rr}}{\hat{g}_{rr}}\right)  \frac{L^{2}}{\tilde{g}_{\phi\phi}}%
-\frac{q_{rr}}{\hat{g}_{rr}}(\dot{z}^{2}+h^{2}G^{2})\right]  \left(
e^{2a}\dot{z}\right)  .
\end{align}
As the right hand side is of the order of $GM,$ according to our
approximation, we can use the zero order solution, eq. (\ref{tmotion}), in the
last term to obtain the following expression:%
\begin{equation}
\frac{d}{d\lambda}\left[  e^{2a}\left(  h^{2}G^{2}\left(  \varphi\right)
+\dot{z}^{2}\right)  \right]  =\left[  -\left(  \frac{q_{tt}}{g_{tt}}%
-\frac{q_{rr}}{\hat{g}_{rr}}\right)  E^{2}+\left(  \frac{q_{\phi\phi}}%
{g_{\phi\phi}}-\frac{q_{rr}}{\hat{g}_{rr}}\right)  \frac{L^{2}}{r^{2}}%
-\frac{q_{rr}}{\hat{g}_{rr}}\dot{z}_{0}^{2}\right]  \dot{z}.
\end{equation}
To proceed further let us make some considerations about the particle's
motion. In the absence of $M$, as we have seen, the transversal motion is
bounded and therefore the particle oscillates around the center of the brane
with a certain characteristic frequency $\omega$. It seems reasonable to admit
that the frequency is very high in order to explain the absence of any
phenomenological trace of extra dimensions in measurements so far. Therefore,
we can assume that during a complete period of oscillation $\left(
\sim1/\omega\right)  $ the relative change of radial coordinate $\left(
\delta r/r\right)  $ is negligible. Under this hypothesis, which is equivalent
to the condition:
\begin{equation}
\frac{\dot{r}}{r}<<\omega, \label{freq}%
\end{equation}
we can integrate directly the above equation to find%
\begin{equation}
e^{2a}G^{2}\dot{z}^{2}=\dot{z}_{0}^{2}-(V_{0}\left(  z\right)  +V_{M}\left(
r,z\right)  ), \label{tmotionMint}%
\end{equation}
where $V_{M}$ is a new term of the effective potential that arises due to
presence of the mass $M$. It is given by%
\begin{equation}
V_{M}\left(  r,z\right)  =-\int_{0}^{z}\left[  -\left(  \frac{q_{tt}}{g_{tt}%
}-\frac{q_{rr}}{\hat{g}_{rr}}\right)  E^{2}+\left(  \frac{q_{\phi\phi}%
}{g_{\phi\phi}}-\frac{q_{rr}}{\hat{g}_{rr}}\right)  \frac{L^{2}}{r^{2}}%
-\frac{q_{rr}}{\hat{g}_{rr}}\dot{z}_{0}^{2}\right]  dz^{\prime}%
\end{equation}
In the first order with respect to $GM$ and for small amplitude oscillation we
obtain:%
\begin{equation}
V_{M}\left(  r,z\right)  =\left(  \frac{2GM\ell^{2}z^{2}}{r^{5}}\right)
\left[  E^{2}+\frac{5}{2}\frac{L^{2}}{r^{2}}+\dot{z}_{0}^{2}\right]
\label{Vm}%
\end{equation}

Therefore, the transversal motion is still bounded for small initial
perturbations $\dot{z}_{0}$. Furthermore, based on (\ref{tmotionMint}), we can
say that the frequency of the oscillation and also the amplitude now depend on
the radial position of the particle. This is the major effect of the mass $M$
on the movement in the extra-dimension direction.

\subsection{Deflection of light}

Let us investigate the influence of the transverse motion over the particle
path in the four-dimensional spacetime. Starting form the equation
(\ref{dispersion}) and reparametrizing it in terms of the angular coordinate
$\phi$, we obtain in the first order of $GM$:%
\begin{align}
&  -E^{2}\left[  1+\frac{2GM}{r}\left(  1+\frac{2\ell^{2}}{3r^{2}}-\frac
{2\ell^{2}}{r^{4}}z^{2}\right)  \right]  +L^{2}r^{2}\left(  1+\frac
{3GM\ell^{2}}{r^{5}}z^{2}\right) \nonumber\\
&  +\left[  1+\frac{2GM}{r}\left(  1+\frac{\ell^{2}}{r^{2}}+\frac{\ell^{2}%
}{r^{4}}z^{2}\right)  \right]  \left(  \frac{dr}{d\phi}\right)  ^{2}%
\frac{L^{2}}{r^{4}}\nonumber\\
&  =-e^{2a}\left(  h^{2}G^{2}+\dot{z}^{2}\right)  . \label{rmotion}%
\end{align}

This equation can be written only in terms of $r$ and $z$ if we eliminate
$\dot{z}^{2}$ by using the equation (\ref{tmotionMint}). Despite this
simplification, it is not easy to solve that equation, since $z$ is evolving
in time. However, as the time scale of the oscillation is much smaller than
the time scale corresponding to the variations of the 4D-motion (see condition
(\ref{freq})), we can replace $z^{2}$ by its average, $\,\sigma^{2}%
=\left\langle z^{2}\right\rangle ,$ taken over several oscillation periods.
Proceeding in this way, we can write equation (\ref{rmotion}) as a function of
the radial coordinate only. Finally, introducing the coordiante $u=1/r$ and
rewriting $\dot{z}_{0}^{2}$ as $\mu^{2}$, the equation assumes, up to
$\sigma^{2}$-order, the following form:%

\begin{equation}
\left(  \frac{du}{d\phi}\right)  ^{2}=F\left(  u\right)  , \label{umotion}%
\end{equation}
where%
\begin{align}
&  F\left(  u\right)  =\left(  \frac{E^{2}-\mu^{2}}{L^{2}}\right)
-u^{2}+\frac{\mu^{2}}{L^{2}}2GMu+2GMu^{3}\left[  1-\left(  \frac{E^{2}%
-3\mu^{2}}{L^{2}}\right)  \frac{\ell^{2}}{3}\right] \nonumber\\
&  +2GMu^{5}\ell^{2}\left[  1+5\left(  \frac{\mu^{2}-E^{2}}{L^{2}}\right)
\sigma^{2}\right]  +10GMu^{7}\ell^{2}\sigma^{2}.
\end{align}
A direct integration of equation (\ref{umotion}) gives us the dependence of
the angular coordinate on the radial position:%
\begin{equation}
\phi\left(  u\right)  -\phi\left(  0\right)  =\int_{0}^{u}\frac{1}%
{\sqrt{F\left(  x\right)  }}dx. \label{phi}%
\end{equation}
To find an explicit solution, it is convenient to note that, if $M=0$, the
particle's trajectory would be a straight line $\left(  u=1/b\sin\phi\right)
,$ where $b=\sqrt{L^{2}/\left(  E^{2}-\mu^{2}\right)  }$ is the impact
parameter. The mass $M$, in the weak field regime $(GMu<<1)$, causes a small
deviation in the particle's original path. Due to the gravitational attraction
of $M$, the minimum separation distance ($r_{0}$) between the particle and the
mass $M$ will be lesser than $b$ by an amount of the order of $GM/b$ in the
first approximation order. For the calculation of the above integral it is
important to determine $r_{0}$ (or equivalently $u_{0}$). As we can see from
the equation (\ref{umotion}), $u_{0}$ is a root of $F\left(  u\right)  $. If
we write $u_{0}=\frac{1}{b}\left(  1+\delta\right)  ,$ then, investigating the
roots of $F\left(  u\right)  ,$ we find, in the first order of $GM$, that:%

\begin{equation}
\delta=\frac{GM}{b}\left(  1-\frac{\mu^{2}}{E^{2}}\right)  ^{-1}\left[
1+\frac{2}{3}\frac{\ell^{2}}{b^{2}}\right]  .
\end{equation}
Therefore the minimum distance to the mass $M$ depends on the photon's energy.
This is a consequence of the fact the particle has a non-null effective mass
$\mu$ due to the oscillation in the transverse direction.

After some algebraic manipulation, we can rewrite $F\left(  u\right)  $ as%
\begin{equation}
F\left(  u\right)  =\left(  u_{0}^{2}-u^{2}\right)  \left(  1-\frac{2}{\left(
u+u_{0}\right)  b}\delta-a_{5}u^{5}-a_{3}u^{3}-a_{1}u\right)  ,
\end{equation}
where%
\begin{align}
a_{5}  &  =10GM\ell^{2}\sigma^{2}\\
a_{3}  &  =2GM\ell^{2}\\
a_{1}  &  =2GM\left[  1+\frac{2}{3}\frac{\ell^{2}}{b^{2}}\left(  \frac{E^{2}%
}{E^{2}-\mu^{2}}\right)  \right]  .
\end{align}
With the help of this expression, the integral (\ref{phi}) can be obtained
directly and, therefore, the deflection angle, $\Delta\phi\equiv2\left\vert
\phi\left(  u_{0}\right)  -\phi\left(  0\right)  \right\vert -\pi,$ can be
calculated. In the first order of $GM$, we find that the deflection of the
light by the mass $M$ in the thick brane is given by:%

\begin{equation}
\Delta\phi=\frac{4GM}{b}\left[  \left(  \frac{1-\mu^{2}/2E^{2}}{1-\mu
^{2}/E^{2}}\right)  +\left(  \frac{1-\mu^{2}/3E^{2}}{1-\mu^{2}/E^{2}}\right)
\frac{\ell^{2}}{b^{2}}+\frac{2}{3}\frac{\ell^{2}\sigma^{2}}{b^{4}}\right]
\label{angle}%
\end{equation}
First we have to emphasize that, if there is no transversal perturbation,
i.e., if $\dot{z}_{0}=0$, then the particle moves strictly in the center of
the brane and the expression above recovers the result found in Ref.
\cite{keeton} for the light deflection in the Garriga-Tanaka thin-brane
geometry. On the other hand, when the particle oscillates along the
extra-dimension direction the deflection angle is modified by two
contributions. The first one is related to the effective mass of the particle.
As the particle becomes massive, the deflection becomes dependent on the
photon's energy. Therefore, photons of distinct frequencies will suffer
different deviations. Thus, the mass $M$, by means of its gravitational
influence, could produce a rainbow if a beam of white light passes near it.
The second effect of the transversal motion on the angle deflection comes from
the last term. As we know $\sigma^{2}$ is related to the amplitude of the
oscillation in the transversal direction and its presence can be explained as
follows. Since the particle oscillates rapidly in the fifth direction then, we
may say that it feels not the geometry of the hypersurface $z=0$ but an
effective four-dimensional geometry, which is described by a kind of an
average metric. Of course, this effective geometry should depend on
$\sigma^{2},$ since it measures how far the particle penetrates into the
extra-dimension direction.

\subsection{Time delay}

The time delay can be studied following a very similar procedure. From the
equation below:
\[
\frac{dr}{dt}=\frac{\dot{\phi}}{\dot{t}}\frac{dr}{d\phi},
\]
we can determine the evolution of the radial coordinate with respect to
$t$-coordinate. By using equations (\ref{E}), (\ref{L}) and (\ref{rmotion}),
we can show that the time delay of a light ray that travels from the point
$r_{1}$ to the point $r_{2}$ and, then, is reflected back to the starting
point $r_{1}$ is given by:%

\begin{align}
\Delta t  &  =\frac{4GM}{1-\mu^{2}/E^{2}}+\frac{6GM}{1-\mu^{2}/E^{2}}\left(
1-\frac{1}{3\left(  1-\mu^{2}/E^{2}\right)  ^{1/2}}\right)  \ln\left(
\frac{4r_{1}r_{2}}{r_{0}^{2}}\right) \nonumber\\
&  +\frac{28GM\ell^{2}}{3r_{0}^{2}}+\frac{8GM\ell^{2}}{3r_{0}^{2}}\frac
{1}{\left(  1-\mu^{2}/E^{2}\right)  ^{1/2}}+\frac{16GM\ell^{2}\sigma^{2}%
}{3r_{0}^{4}}\nonumber\\
&  +2\left(  r_{1}+r_{2}\right)  \left(  \frac{\mu^{2}/E^{2}}{1-\mu^{2}/E^{2}%
}\right)  -\frac{2GM}{r_{0}}\frac{\left(  r_{1}+r_{2}\right)  }{\left(
1-\mu^{2}/E^{2}\right)  }\left(  \frac{\mu^{2}/E^{2}}{1-\mu^{2}/E^{2}}\right)
\nonumber\\
&  -\frac{4GM\ell^{2}}{3r_{0}^{2}}\left(  r_{1}+r_{2}\right)  \left(
\frac{\mu^{2}/E^{2}}{1-\mu^{2}/E^{2}}\right)  , \label{delay}%
\end{align}
where $r_{0}$ is the distance of the closest approach to the mass $M$. As it
happens in the deflection of the light ray, the thickness has two major
effects on the flight time of the light ray. As we can see from (\ref{delay}),
the time delay depends on the energy of the light ray and depends also on how
far the particle penetrates in the fifth direction.

\section{Semi-classical approach}

We can approach the problem of the particle motion in the thick brane from a
semi-classical point of view. The idea is to consider the quantization of the
motion equation in the $z-$direction, while the movement in the 4D-world will
be still described classically. Substituting $\dot{z}$ by the operator
$i\partial_{z}$ (applying naively the usual quantization rules), the equation
assumes a similar form of the Schroedinger equation:%
\begin{equation}
-e^{2a}\frac{\partial^{2}\psi}{\partial z^{2}}+V_{0}\psi=\mu^{2}\psi.
\end{equation}
In this picture, the parameter $\mu$ is interpreted as the eigenvalue of the
operator that appears in the left hand side of the equation. This operator can
be viewed as a kind of mass operator since the eingenvalues $\mu$ plays the
role of the effective mass of the particle when its motion in the 4D-world is
considered. The mass operator is hermitian if the warping factor $e^{-2a}$ is
taken as a weight function for the inner product. Choosing $G\left(
\varphi\right)  $ appropriately, then, we can find a bound state with
eingenvalue $\mu=0$, the so-called zero-mode state.

When the body of mass $M$ is taken into account, the situation changes a bit
because of the presence of the potential $V_{M}$. However, we can treat
$V_{M}$, which depends on $r$-coordinate, as a small pertubation of the main
operator. Of course, $r$ is evolving in time, but according to the condition
(\ref{freq}), this variation can be considered as an adiabatic process.
Therefore, we can find corrections to the eigenfunctions and eigenvalues by
using the standard procedure of perturbation method in quantum mechanics. The
new ground state $\psi_{0}^{M}$, for instance, which originally was the
zero-mode $\psi_{0}$, is now associated to the non-zero eigenvalue
$\left\langle \psi_{0}\right\vert V_{M}\left\vert \psi_{0}\right\rangle $,
i.e., the expected value of the potencial $V_{M}$ evaluated in the unperturbed
zero-mode state $\psi_{0}$. Considering equation (\ref{Vm}), we find%
\begin{equation}
\left\langle \psi_{0}\right\vert V_{M}\left\vert \psi_{0}\right\rangle
=\left(  \frac{2GM\ell^{2}\sigma^{2}}{r^{5}}\right)  \left[  E^{2}+\frac{5}%
{2}\frac{L^{2}}{r^{2}}\right]  ,
\end{equation}
where $\sigma^{2}=\left\langle \psi_{0}\right\vert z^{2}\left\vert \psi
_{0}\right\rangle $ measures now the width in the transverse direction of the
wave function corresponding to the zero-mode.

On the other hand, the equations of motion in the 4D-world, as the equation
(\ref{rmotion}), depend on $z$ and $\dot{z}$, which should be considered as
operators in our scheme. Therefore, in order to describe the motion
classically, we have to take the average of the equation in a certain quantum
state that describes the particle's state with respect to the fifth direction.
In particular, the equation of the light deflection for the light ray in the
fundamental state $\psi_{0}^{M}$ is given by:
\begin{align}
&  -E^{2}\left(  1+\frac{2GM}{r}\left(  1+\frac{2\ell^{2}}{3r^{2}}-\frac
{2\ell^{2}}{r^{4}}\sigma^{2}\right)  \right)  +L^{2}r^{2}\left(
1+\frac{3GM\ell^{2}}{r^{5}}\sigma^{2}\right) \nonumber\\
&  +\left(  1+\frac{2GM}{r}\left(  1+\frac{\ell^{2}}{r^{2}}+\frac{\ell^{2}%
}{r^{4}}\sigma^{2}\right)  \right)  \left(  \frac{dr}{d\phi}\right)  ^{2}%
\frac{L^{2}}{r^{4}}\nonumber\\
&  =\left(  \frac{2GM\ell^{2}\sigma^{2}}{r^{5}}\right)  \left[  E^{2}+\frac
{5}{2}\frac{L^{2}}{r^{2}}\right]  .
\end{align}
As the brane has a non-null thickness, then $\sigma$ is not zero. Following
the same steps previously described, we can obtain the deviation angle for the
light ray in the zero-mode:%

\begin{equation}
\Delta\phi=\frac{4GM}{b}\left[  1+\frac{\ell^{2}}{b^{2}}+\frac{2}{3}\frac
{\ell^{2}\sigma^{2}}{b^{4}}\right]  .
\end{equation}
In the zero-mode, the particle has a null mass, so the effect of the thickness
of the brane is now exclusively encoded in the width of the wave function with
respect to the extra-dimensional direction. Following a similar procedure, we
can show that the time delay of light rays in the zero-mode is given by
equation (\ref{delay}) with $\mu=0$.

The data about the deflection of light in the solar system \cite{dados}
imposes an upper bound on $\ell$. Roughly, we have $\ell<10^{4}Km$. On its
turn, the second correction is $\sigma^{2}/b^{2}$ times lesser than $\ell
^{2}/b^{2}$. If we consider, as reference values for $\varepsilon$ and $b$,
the thickness of a TeV-brane and the Solar radius $R_{\odot}$ respectively,
then we can write:
\begin{equation}
\frac{\sigma}{b}\sim10^{-25}\left(  \frac{\sigma}{\varepsilon}\right)  \left(
\frac{R_{\odot}}{b}\right)  .
\end{equation}
Therefore, the influence of the brane thickness over the bending of the light
is very tiny in the solar system. Nevertheless, it is reasonable to expected
that the effects become more significant for microscopic black holes in the
strong field regime. However, this situation cannot be considered within our scheme.

\section{Final remarks}

Previous works have investigated classical tests of General Relativity in
braneworld scenario, trying to find empiric constraints on parameters of
infinitely thin brane models. Here, we have analyzed the deflection of light
and the time delay in the context of a thick brane scenario in order to
determine the effects of the brane thickness over the motion in the 4D-world.
The thick brane here is treated as a self-gravitating domain wall that
corresponds to a regularized version of a infinitely thin brane in the RSII
model. Considering a confined mass $M$ in the thick brane, we find an
approximate solution of this configuration, propagating initial data from the
center of the brane to the fifth direction by using the Einstein equations
coupled to the scalar field. The solution is built from the Garriga-Tanaka
metric, taken as part of the initial data. Hence we may say that it represents
approximately the gravitational field of a black hole in the weak field limit
and in the vicinity of the brane center.

Based on a mechanism that describes the confinement of massive test particles
in a domain wall, by means of a direct interaction between the particle and
the scalar field, we developed a formalism to deal with the motion of massless
particles in a thick brane. According to this prescription the particle gains
an effective mass when the motion has a transversal component. The variation
of the mass with respect to the fifth direction is dictated by a certain
function $G\left(  \varphi\right)  $ of the scalar field. Bound motions are
found if $G\left(  \varphi\right)  $ is conveniently chosen.

With the help of this formalism, we have studied the motion of massless
particles in the spacetime produced by a mass $M$ confined in an
self-gravitating thick brane. The metric in the bulk is not warped and,
therefore, the transverse motion is not decoupled from the movement in the
radial direction. However, if the motion in the transversal direction is a
high frequency oscillation, then we can find an approximate solution for the
light deflection. We have shown that the transverse motion influences the
bending of the light rays in two different ways. The first one is related to
the effective mass the particle acquires due to its motion in the fifth
dimension. As a consequence, the deviation angle becomes dependent on the
light ray energy and, because of this feature, the mass \ $M$ may produce the
interesting phenomenon of gravitational rainbow. The second effect is related
to the fact that, as the particle oscillates rapidly in the fifth direction,
it sees a kind of effective four-dimensional geometry. It follows, therefore,
that the deviation angle shows a dependence on the amplitude of the
transversal motion.

We have also considered a semi-classical approach to the problem. Analyzing
the quantization of the motion in the fifth direction, we have seen that the
system may admit a zero-mode solution, if $G\left(  \varphi\right)  $ has
appropriate features. Based on this solution, we have studied the motion of
the zero-mode in the 4D-world at classical level. Specifically, we have
determined the effects of the thickness of the brane over the deflection of
the light rays in the zero-mode by a mass $M$. Compared to the thin brane
result, the deviation angle has an additional contribution, which depends on
the width, or more precisely, the root-mean-square deviation $\left(
\sigma\right)  $, of the zero-mode in the extra-dimension direction. This
additional term is a consequence of the fact that in thick branes the
confinement of particles is not a delta-like confinement in a hypersurface,
once the wave function has non-null width in the extra-dimensional direction.
Thus, we may conclude that the result suggest that, regarding its apparent
motion in the 4D-world, the particle does not feel the geometry of the
hypersurface $z=$ 0, but an effective four-dimensional geometry that depends
on the profile of the wave function in the fifth direction.

\end{document}